\documentclass[preprint,showpacs,preprintnumbers,amsmath,amssymb]{revtex4}
\usepackage{graphicx}
\usepackage{dcolumn}
\usepackage{bm}
\usepackage{color}
\usepackage{hhline}

\begin{document}

\title{Simulation of spin-polarized scanning tunneling microscopy on complex magnetic surfaces:
       Case of a Cr monolayer on Ag(111)}

\author{Kriszti\'an Palot\'as}
\email{palotas@phy.bme.hu}
\affiliation{Budapest University of Technology and Economics,
Department of Theoretical Physics, Budafoki \'ut 8., H-1111 Budapest, Hungary}

\author{Werner A. Hofer}
\affiliation{University of Liverpool, Surface Science Research Centre, L69 3BX Liverpool, UK}

\author{L\'aszl\'o Szunyogh}
\affiliation{Budapest University of Technology and Economics,
Department of Theoretical Physics, Budafoki \'ut 8., H-1111 Budapest, Hungary }

\date{\today}

\begin{abstract}

We propose an atom-superposition-based method for simulating spin-polarized scanning tunneling microscopy (SP-STM)
from first principles. Our approach provides bias dependent STM images in high spatial resolution, with the capability of using
either constant current or constant height modes of STM. In addition, topographic and magnetic contributions can clearly be
distinguished, which are directly comparable to results of SP-STM experiments in the differential magnetic mode.
Advantages of the proposed method are that it is computationally cheap, it is easy to parallelize, and it can employ the results
of any ab initio electronic structure code.
Its capabilities are illustrated for the prototype frustrated hexagonal antiferromagnetic system, Cr monolayer on Ag(111) in a
noncollinear magnetic $120^{\circ}$ N\'eel state. We show evidence that the magnetic contrast is sensitive to the
tip electronic structure, and this contrast can be reversed depending on the bias voltage.

\end{abstract}


\maketitle

\section{Introduction}

Research on magnetic systems has intensified in the last decade with the ultimate aim to produce magnetic data storage devices
with ultrahigh information density \cite{ultrahigh,weiss05}. This can be achieved by reducing the size of the information storage
units going down to the nanoscale or even to single atoms \cite{serrate10}. Detecting and manipulating spins \cite{tao09}
with high accuracy on the atomic scale is, thus, essential for future technological applications.
Spin-polarized scanning tunneling microscopy (SP-STM) \cite{bode03review} is one of the main tools for studying magnetism at the
atomic scale. Recent experimental advances using this technique allow the investigation of complex magnetic structures (frustrated
antiferromagnets, spin spirals, etc.) \cite{bode07,ferriani08,gao08,wulfhekel10review,wiesendanger09review,wasniowska10}.
Considering such structures in reduced dimensions, their magnetic ground state can be determined \cite{ferriani08,wasniowska10}
and the nature of magnetic interactions can be studied by theoretical means, e.g. from first principles \cite{antal08}, or
applying a multiscale approach \cite{udvardi08}. However, a proper validation of the proposed ground state spin structures demands
a method, which is capable to directly compare them to experimental observations. This can be done by SP-STM simulations.

Our motivation was to construct a computationally cheap and user-friendly, yet reliable model for simulating SP-STM.
Here, we propose an efficient method based on the spin-polarized Tersoff-Hamann model \cite{wortmann01} and the atom superposition
approach \cite{heinze06,yang02,smith04}. Our model goes beyond the work of Heinze \cite{heinze06}, and considers in particular,\\
(1) bias voltage, for simulating bias dependent physical properties,\\
(2) energy dependence of the vacuum decay of electron states, and\\
(3) energy dependence of atomic local spin quantization axes.\\
The main advance of our tunneling model is the inclusion of the tip electronic structure, which is neglected in
Refs.\ \cite{wortmann01,heinze06}, and it enables to study tip effects on the SP-STM images.

First, we determine energy dependent virtual differential conductance ($dI/dU$) quantities on a three-dimensional fine real space
grid from electron local density of states ($LDOS$).
Integrating the differential current ($dI$) contributions in an energy window in accordance with the applied bias voltage, we
obtain a three-dimensional current map on the same grid, from which constant current or constant height images can be extracted.
Calculating differential current first and then tunneling current by integration over energies proved to be numerically more
stable than the opposite (numerical differentiation of tunneling current with respect to energy) \cite{hofer05sts}.
Furthermore, bias dependent apparent barrier height can be included in our approach.

The paper is organized as follows: The theoretical model of SP-STM is presented in section \ref{sec_spstm}, where we explicitly
point out extensions to the Heinze model \cite{heinze06} and other atom-superposition-based approaches \cite{yang02,smith04}.
As an application, we investigate the frustrated hexagonal antiferromagnet, one monolayer (ML) Cr on Ag(111)
in section \ref{sec_res}.
We study two magnetic chiralities for its noncollinear N\'eel state and determine the energetically favored magnetic structure.
Moreover, we compare electronic structures of Cr obtained from collinear and noncollinear calculations.
By simulating SP-STM images we are able to investigate magnetic contrast depending on tip electronic structure and bias voltage.
Our conclusions are found in section \ref{sec_conc}.

\section{Theoretical model of SP-STM}
\label{sec_spstm}

In SP-STM the total tunneling current can be written as a sum of a non-spin-polarized, $I_0$, and a spin-polarized part, $I_P$
\cite{wortmann01,hofer03,smith04,heinze06},
\begin{equation}
\label{Eq_Current_decomp}
I=I_0+I_P.
\end{equation}
This formula is generally valid for either collinear or noncollinear surface and tip spin structures.
$I_0$ and $I_P$ can be calculated at different levels of approximation for the tunneling current.
It could, in principle, be implemented within the multiple scattering framework \cite{palotas05}. While the perturbation approach
has been used by Hofer and Fisher \cite{hofer03} for collinear surface and tip spin structures with an arbitrary angle between
their spin quantization axes, the most commonly used method is based on the Tersoff-Hamann model \cite{tersoff83,tersoff85}.
Wortmann et al.\ \cite{wortmann01} introduced its spin-polarized version applicable to complex noncollinear surface spin
structures. Heinze \cite{heinze06} combined this with the atom superposition method \cite{yang02,smith04}. Note that in the
following we denote $I_0$ and $I_P$ by $I_{TOPO}$ (topographic current) and $I_{MAGN}$ (magnetic current), respectively.

Here, based on the work of Heinze \cite{heinze06}, we propose a hybrid model, which uses essentially the Tersoff-Hamann
formalism but we do not restrict the tip electron density of states (DOS) to be constant in energy. This means that different tip
models \cite{ferriani10tip} and their effect on tunneling properties can be investigated. The only requirement for our present
formalism is that we assume that electrons tunnel through one tip apex atom. Since the tip electronic structure is explicitly
included in our method via the projected DOS onto this apex atom, Eq.(2) of Ref.\ \cite{heinze06} needs to be reconsidered.
Our strategy is to determine differential currents first and then perform an energy integral in a window according to
the applied bias voltage ($V$) in order to arrive at the tunneling current.

Let us define the following position- and energy-dependent density matrices in spin space for tip (T) and sample (S), respectively,
\begin{eqnarray}
\label{Eq_rho}
\underline{\underline{\rho}}_{T,S}(\overline{r},E)&=&n_{T,S}(\overline{r},E)\underline{\underline{I}}+\overline{m}_{T,S}(\overline{r},E)\underline{\underline{\overline{\sigma}}}\\
&=&
\begin{bmatrix}
n_{T,S}(\overline{r},E)+m_{T,S}^z(\overline{r},E)    & m_{T,S}^x(\overline{r},E)-im_{T,S}^y(\overline{r},E) \\
m_{T,S}^x(\overline{r},E)+im_{T,S}^y(\overline{r},E) & n_{T,S}(\overline{r},E)-m_{T,S}^z(\overline{r},E)
\end{bmatrix}.\nonumber
\end{eqnarray}
Here, $\underline{\underline{I}}$ is the $2\times 2$ unit matrix, $\underline{\underline{\overline{\sigma}}}$ is the Pauli matrix
vector, while $n_{T}(\overline{R}_{TIP},E)$ and $\overline{m}_{T}(\overline{R}_{TIP},E)$ are the charge and magnetization DOS
projected to the tip apex atom. On the other hand, $n_{S}(\overline{r},E)$ and $\overline{m}_{S}(\overline{r},E)$ are the charge
and magnetization DOS of the sample surface at position $\overline{r}$.
They can be obtained from the corresponding density matrix as
\begin{eqnarray}
\label{Eq_n_rho}
n_{T,S}(\overline{r},E)=\frac{1}{2}Tr\left(\underline{\underline{\rho}}_{T,S}(\overline{r},E)\right),\\
\overline{m}_{T,S}(\overline{r},E)=\frac{1}{2}Tr\left(\underline{\underline{\rho}}_{T,S}(\overline{r},E)\underline{\underline{\overline{\sigma}}}\right),
\label{Eq_m_rho}
\end{eqnarray}
where the Trace is performed in the spin space.
At the tip position $\overline{R}_{TIP}$ above the surface we obtain the charge and magnetization electron local density of states
($LDOS$) of the surface in vacuum, $n_{S}(\overline{R}_{TIP},E)$ and $\overline{m}_{S}(\overline{R}_{TIP},E)$, respectively.
Combining the sample and tip density matrices at $\overline{R}_{TIP}$, a modified $LDOS$ can be defined as
\begin{eqnarray}
\label{Eq_rho_rho}
LDOS(\overline{R}_{TIP},E)&=&\Delta E\frac{1}{2}Tr\left(\underline{\underline{\rho}}_S(\overline{R}_{TIP},E)\underline{\underline{\rho}}_T(\overline{R}_{TIP},E)\right)\\
&=&\Delta E\left(n_S(\overline{R}_{TIP},E)n_T(\overline{R}_{TIP},E)+\overline{m}_S(\overline{R}_{TIP},E)\overline{m}_T(\overline{R}_{TIP},E)\right),\nonumber
\end{eqnarray}
which, in fact, combines the vacuum $LDOS$ of the surface and the projected DOS of the tip apex atom.
This formula is consistent with the spin-polarized Tersoff-Hamann model \cite{wortmann01}, except the fact that it explicitly
includes the electronic structure of the tip apex.
Here, $\Delta E$ ensures that the $LDOS$ is correctly measured in units of $(eV)^{-1}$.
Note that in Ref.\ \cite{heinze06} the tunneling current was proportional to a dimensionless $LDOS$ at the Fermi level.

The vacuum $LDOS$ of the surface can be approximated by a superposition of decaying atomic electron states.
Following this, we consider the position dependence of the sample density matrices as
$\underline{\underline{\rho}}_S(\overline{R}_{\alpha},E)$ with $\overline{R}_{\alpha}$ the position vector of the $\alpha$th
sample surface atom, in order to allow different chemical or magnetic properties for these atoms.
$n_{S}^{\alpha}(E)=n_{S}(\overline{R}_{\alpha},E)$ and $\overline{m}_{S}^{\alpha}(E)=\overline{m}_{S}(\overline{R}_{\alpha},E)$
now denote charge and magnetization DOS projected to the $\alpha$th surface atom, respectively.
It has to be noted that chemical differences between surface atoms were not taken into account in Ref.\ \cite{heinze06}.
A tunneling transition between the tip apex and the $\alpha$th surface atom at energy $E$ can be represented as the Trace of the
multiplied density matrices, similarly to Eq.(\ref{Eq_rho_rho}). This is the energetic ingredient for the tunneling transition.
Apart from this, the transmission coefficient through a potential barrier between the $\alpha$th
surface atom and the tip apex has to be included in the tunneling model, in general we denote it by $T(E,V,d_{\alpha})$.
It has energy and bias dependence, and contains geometry information of the three-dimensional tunnel junction via the
distance between the tip apex and the $\alpha$th surface atom,
\begin{equation}
\label{Eq_d_alpha}
d_{\alpha}(x,y,z)=\left|\overline{R}_{TIP}(x,y,z)-\overline{R}_{\alpha}\right|.
\end{equation}
Thus, the modified $LDOS$ at the tip apex position $\overline{R}_{TIP}(x,y,z)$ and at energy $E$ can be approximated as the
superposition of individual atomic contributions from the sample surface as
\begin{equation}
\label{Eq_LDOS_general}
LDOS(x,y,z,E,V)=\Delta E\sum_{\alpha}T(E,V,d_{\alpha}(x,y,z))\frac{1}{2}Tr\left(\underline{\underline{\rho}}_S(\overline{R}_{\alpha},E)\underline{\underline{\rho}}_T(\overline{R}_{TIP},E)\right).
\end{equation}
The main advantage of using the density matrix formalism is that electronic and spin structures calculated either
nonrelativistically or relativistically can be treated within the same theoretical framework. 

We calculate the above $LDOS$ values at $(x,y,z)$ grid points of a three-dimensional fine grid in a finite box
above the surface. The image resolution is determined by the density of $(x,y)$ grid points.
The motivation for using the atomic superposition approximation is, on one hand, computational efficiency, since calculating and
storing the projected DOS onto surface atoms in the magnetic unit cell is computationally much cheaper compared to the vacuum
$LDOS$ of the surface on a great number of grid points. On the other hand, such atom-projected DOS functions are routinely obtained
in all ab initio electronic structure codes, whereas vacuum $LDOS$ is not always routinely accessible for the average user.

According to above, the $LDOS$ can be decomposed, similarly to Eq.(\ref{Eq_Current_decomp}), as
\begin{equation}
\label{Eq_LDOS_decomp}
LDOS(x,y,z,E,V)=LDOS_{TOPO}(x,y,z,E,V)+LDOS_{MAGN}(x,y,z,E,V),
\end{equation}
and assuming an exponential decay of the electron wavefunctions, the TOPO and MAGN terms can be written as
\begin{eqnarray}
\label{Eq_LDOSTOPO}
LDOS_{TOPO}(x,y,z,E,V)&=&\Delta E\sum_{\alpha}e^{-2\kappa(E,V)d_{\alpha}(x,y,z)}n_T(E)n_S^{\alpha}(E)\\
LDOS_{MAGN}(x,y,z,E,V)&=&\Delta E\sum_{\alpha}e^{-2\kappa(E,V)d_{\alpha}(x,y,z)}m_T(E)m_S^{\alpha}(E)cos\varphi_{\alpha}(E).
\label{Eq_LDOSMAGN_coll}
\end{eqnarray}
Here the sum over $\alpha$ has to be carried out, in principle, over all the surface atoms. Convergence tests, however,
showed that including a relatively small number of atoms in the sum provides converged $LDOS$ values \cite{palotas11sts}.
Each surface atom is characterized by a local spin quantization axis, $\overline{e}_S^{\alpha}(E)$, which can be defined
from the sample magnetization DOS vector as $\overline{e}_S^{\alpha}(E)=\overline{m}_S^{\alpha}(E)/|\overline{m}_S^{\alpha}(E)|$.
In the most general case, these local axes can be energy dependent, see Eqs.(\ref{Eq_MDOS_noncoll}) and (\ref{Eq_axis}),
and Table \ref{Table_axis}. This also holds for the spin quantization axis of the tip apex, $\overline{e}_T(E)$.
The exponential factor is the transmission probability for electrons tunneling between states of atom $\alpha$ on the surface and
the tip apex, where $\kappa$ is the vacuum decay. $\kappa$ is treated within the independent-orbital approximation
\cite{tersoff83,tersoff85,heinze06}, which means that the same (spherical) decay is used for all type of orbitals,
but its energy dependence is explicitly considered essentially in the same fashion as in Ref.\ \cite{lang86}.
Extension of our model to take into account orbital dependent vacuum decay following Chen's work \cite{chen90} is planned
in the future, which is relevant for a more advanced description of tunneling from/to directional orbitals.
In the present paper we consider two different ways of calculating $\kappa$. One is inspired by the Tersoff-Hamann model,
taking only surface properties into account,
\begin{equation}
\label{Eq_kappa_TH}
\kappa(E)=\frac{1}{\hbar}\sqrt{2m(\phi_S-|E-E_F^S|)},
\end{equation}
where the electron's mass is $m$, $\hbar$ is the reduced Planck constant,
while $\phi_S$ and $E_F^S$ are the average electron workfunction and the Fermi energy of the sample surface, respectively.
The absolute value ensures that the transmission probability is symmetric in the positive and negative bias range with the
minimum at zero bias \cite{becker10_prb,becker10_apl}. This way we circumvent the problem of the bias-asymmetric contribution
from the sample $LDOS$ to the differential conductance \cite{ukraintsev96}.
We use this energy dependent vacuum decay for an ideal, electronically featureless and maximally spin-polarized tip model.
Note that this formula does not have an explicit bias dependence.
Taking into account the tip apex electronic structure obtained from first principles, the more general expression for
$\kappa$ is based on the one-dimensional Wentzel-Kramers-Brillouin (WKB) approximation assuming an effective rectangular
potential barrier between the tip and the surface,
\begin{equation}
\kappa(E,V)=\frac{1}{\hbar}\sqrt{2m\left(\frac{\phi_S+\phi_T+eV}{2}-|E-E_F^S|\right)},
\label{Eq_kappa_WKB}
\end{equation}
with $\phi_T$ being the local electron workfunction of the tip apex, $e$ the elementary charge and $V$ the applied bias voltage.
This vacuum decay formula is considered for our magnetic Ni tip model.
The quantity $(\phi_S+\phi_T+eV)/2-|E-E_F^S|$ is the energy- and bias dependent apparent barrier height for tunneling electrons,
$\phi_a(E,V)$. Empirical or model nonlinear variations of $\phi_a(E,V)$ with respect to bias voltage
\cite{becker10_prb,becker10_apl} can also be included in our approach.
Note that in the case of $\phi_T+eV=\phi_S$, the first expression of $\kappa$, Eq.(\ref{Eq_kappa_TH}) is recovered.
The average workfunction of the sample surface is calculated from the local electrostatic potential on a
three-dimensional fine grid, $\Phi(x,y,z)$, as
\begin{equation}
\label{Eq_WF_average}
\phi_S=\max_z\left\{\frac{1}{N_xN_y}\sum\limits_{x,y}\Phi(x,y,z)\right\}-E_F^S,
\end{equation}
with $N_x$ and $N_y$ the corresponding number of grid points,
and the local workfunction of the tip apex is obtained as
\begin{equation}
\label{Eq_WF_local}
\phi_T=\max_z\left\{\Phi(x_0,y_0,z)\right\}-E_F^T,
\end{equation}
with $x_0$ and $y_0$ lateral coordinates of the tip apex atom, and $E_F^T$ the Fermi energy of the tip material.

In the $LDOS$ formula Eq.(\ref{Eq_LDOSMAGN_coll}), $\varphi_{\alpha}(E)$ is the angle between the spin quantization axes of the
tip apex and the $\alpha$th surface atom at energy $E$.
Previously, only the case of energy independent $\varphi_{\alpha}$ has been considered \cite{heinze06}, which corresponds to the
angle between the directions of local magnetic moments of surface atoms and the tip magnetic moment.
However, there are more possibilities to combine electronic structure data of sample and tip, which may result in an
energy dependent $\varphi_{\alpha}(E)$, see Table \ref{Table_axis}.
All listed combinations can be investigated within our formalism. The combinations framed by black solid lines are considered in
the present work, while the one with the gray dashed line corresponds to the studied system in Ref.\ \cite{palotas11sts}.
In Eq.(\ref{Eq_LDOSMAGN_coll}), $m_T(E)$ and $m_S^{\alpha}(E)$ denote electron magnetization DOS projected to the
tip apex and the $\alpha$th surface atom, respectively, in the collinear case,
\begin{equation}
\label{Eq_MDOS_coll}
m_{T,S}(E)=n_{T,S}^{\uparrow}(E)-n_{T,S}^{\downarrow}(E),
\end{equation}
$\uparrow$ and $\downarrow$ relative to their local spin quantization axes.
Similarly, in Eq.(\ref{Eq_LDOSTOPO}), $n_T(E)$ and $n_S^{\alpha}(E)$ are electron charge DOS projected to the tip apex
and the $\alpha$th surface atom, respectively,
\begin{equation}
\label{Eq_CDOS_coll}
n_{T,S}(E)=n_{T,S}^{\uparrow}(E)+n_{T,S}^{\downarrow}(E).
\end{equation}
The spin-resolved atom-projected DOS ($PDOS$) quantities, $n_{T,S}^{\uparrow,\downarrow}(E)$,
are obtained from first principles collinear magnetic calculations. For this task any available ab initio
electronic structure code can be used. This flexibility of the present SP-STM approach is expected to be highly advantageous.
Spin-resolved $PDOS$ is considered by assuming a Gaussian broadening of the peaks at the
k-resolved spin-dependent electron energy (Kohn-Sham) eigenvalues, $\varepsilon_{T,S}^{j\uparrow,\downarrow}(\overline{k})$,
obtained at zero temperature, as
\begin{equation}
\label{Eq_PDOS}
n_{T,S}^{\uparrow,\downarrow}(E)=\sum_{\overline{k}}\sum_j\frac{1}{G\sqrt{\pi}}e^{-\left(E-\varepsilon_{T,S}^{j\uparrow,\downarrow}(\overline{k})\right)^2/G^2}\int\limits_{atomic\;volume}d^3r\psi_{T,S}^{j\overline{k}\uparrow,\downarrow\dagger}(\overline{r})\psi_{T,S}^{j\overline{k}\uparrow,\downarrow}(\overline{r}),
\end{equation}
with $\psi_{T,S}^{j\overline{k}\uparrow,\downarrow}(\overline{r})$ the spin-dependent electron wavefunctions corresponding to
$\varepsilon_{T,S}^{j\uparrow,\downarrow}(\overline{k})$ for tip (T) and surface (S), respectively, and $j$ the energy band index.
The integral over the atomic volumes can be performed either in the atomic sphere or within the Bader volume \cite{tang09}.
In the present study we use integral over atomic spheres.
The Gaussian parameter $G$ could, in general, be temperature dependent. In our calculations, we fixed it to a relatively
high value of 0.1 eV in order to provide smooth $n_{T,S}^{\uparrow,\downarrow}(E)$ functions.
Concerning smoothness of $PDOS$, a high $G$ value counteracts the effect of eventually underrepresented bulk states due to
a slab geometry, and it is useful if the number of k-points in the Brillouin zone is restricted due to computational reasons.

As Heinze pointed out \cite{heinze06}, in case of having chemically equivalent surface atoms, the spin structure plays a
much more dominant role compared to the detailed electronic structure in determining the main features of an SP-STM image.
This means that SP-STM simulation of a known noncollinear spin structure can reasonably be approximated based on the
collinear electronic structure. We would like to check this statement, therefore, in this paragraph we show how to incorporate
the fully noncollinear electronic structure into our model.
In this case the atom-projected charge DOS at energy $E$ is obtained in the following way,
\begin{equation}
\label{Eq_CDOS_noncoll}
n_{T,S}(E)=\sum_{\overline{k}}\sum_j\frac{1}{G\sqrt{\pi}}e^{-\left(E-\varepsilon_{T,S}^{j}(\overline{k})\right)^2/G^2}\int\limits_{atomic\;volume}d^3r\Psi_{T,S}^{j\overline{k}\dagger}(\overline{r})\Psi_{T,S}^{j\overline{k}}(\overline{r}),
\end{equation}
where $\varepsilon_{T,S}^{j}(\overline{k})$ is the set of electron energy (Kohn-Sham) eigenvalues at zero temperature, and
$\Psi_{T,S}^{j\overline{k}}(\overline{r})$ the corresponding spinor electron wavefunctions.
The atom-projected magnetization DOS vector at energy $E$ reads
\begin{equation}
\label{Eq_MDOS_noncoll}
\overline{m}_{T,S}(E)=\sum_{\overline{k}}\sum_j\frac{1}{G\sqrt{\pi}}e^{-\left(E-\varepsilon_{T,S}^{j}(\overline{k})\right)^2/G^2}\int\limits_{atomic\;volume}d^3r\Psi_{T,S}^{j\overline{k}\dagger}(\overline{r})\overline{\sigma}\Psi_{T,S}^{j\overline{k}}(\overline{r}),
\end{equation}
with $\overline{\sigma}$ being the Pauli spin operator vector.
Unit vectors determining the local spin quantization axis of the tip apex, $\overline{e}_T(E)$, and the
$\alpha$th surface atom, $\overline{e}_S^{\alpha}(E)$, at a given energy can be calculated as
\begin{equation}
\label{Eq_axis}
\overline{e}_{T,S}(E)=\frac{\overline{m}_{T,S}(E)}{m_{T,S}(E)}=\frac{\overline{m}_{T,S}(E)}{\sqrt{m_{T,S}^x(E)^2+m_{T,S}^y(E)^2+m_{T,S}^z(E)^2}},
\end{equation}
thus, the atom-projected magnetization DOS vector can be rewritten as
\begin{equation}
\label{Eq_MDOS_noncoll_rewrite}
\overline{m}_{T,S}(E)=m_{T,S}(E)\overline{e}_{T,S}(E)=\sqrt{m_{T,S}^x(E)^2+m_{T,S}^y(E)^2+m_{T,S}^z(E)^2}\overline{e}_{T,S}(E).
\end{equation}
Using this expression and the scalar product of the local spin quantization axes,
$\overline{e}_T(E)\overline{e}_S^{\alpha}(E)=cos\varphi_{\alpha}(E)$,
the following holds,
\begin{equation}
\label{Eq_MDOS_noncoll_coll}
\overline{m}_T(E)\overline{m}_S^{\alpha}(E)=m_T(E)\overline{e}_T(E)m_S^{\alpha}(E)\overline{e}_S^{\alpha}(E)=m_T(E)m_S^{\alpha}(E)cos\varphi_{\alpha}(E).
\end{equation}

The $LDOS$ can also be written in terms of energy dependent spin polarizations. The spin polarization is defined as
\begin{equation}
\label{Eq_Spinpol_coll}
P_{T,S}(E)=\frac{m_{T,S}(E)}{n_{T,S}(E)}=\frac{n_{T,S}^{\uparrow}(E)-n_{T,S}^{\downarrow}(E)}{n_{T,S}^{\uparrow}(E)+n_{T,S}^{\downarrow}(E)},
\end{equation}
assuming collinear electronic structure. From noncollinear electronic structure the energy dependent spin polarization vectors
are obtained by using Eq.(\ref{Eq_MDOS_noncoll_rewrite}) as
\begin{equation}
\label{Eq_Spinpol_noncoll}
\overline{P}_{T,S}(E)=\frac{\overline{m}_{T,S}(E)}{n_{T,S}(E)}=\frac{\sqrt{m_{T,S}^x(E)^2+m_{T,S}^y(E)^2+m_{T,S}^z(E)^2}}{n_{T,S}(E)}\overline{e}_{T,S}(E)=P_{T,S}(E)\overline{e}_{T,S}(E).
\end{equation}
The relation between spin polarization vectors and scalars is similar to Eq.(\ref{Eq_MDOS_noncoll_coll}),
\begin{equation}
\label{Eq_Spinpol_noncoll_coll}
\overline{P}_T(E)\overline{P}_S^{\alpha}(E)=P_T(E)\overline{e}_T(E)P_S^{\alpha}(E)\overline{e}_S^{\alpha}(E)=P_T(E)P_S^{\alpha}(E)cos\varphi_{\alpha}(E),
\end{equation}
with the same energy dependent unit vectors, which define the local spin quantization axes, see Eq.(\ref{Eq_axis}).
Thus, the $LDOS$ at the tip apex position and at energy $E$ can alternatively be written using the above defined
spin polarizations as
\begin{equation}
\label{Eq_LDOS_Spinpol}
LDOS(x,y,z,E,V)=\Delta E\sum_{\alpha}e^{-2\kappa(E,V)d_{\alpha}(x,y,z)}n_T(E)n_S^{\alpha}(E)[1+P_T(E)P_S^{\alpha}(E)cos\varphi_{\alpha}(E)].
\end{equation}

Using Eq.(11) of Ref.\ \cite{wortmann01} and our $LDOS$ expression, a virtual differential conductance
at the tip apex position and at energy $E$ can be defined as
\begin{equation}
\label{Eq_dIdV}
\frac{dI}{dU}(x,y,z,E,V)=\frac{e^2}{h}\sum_{\alpha}e^{-2\kappa(E,V)d_{\alpha}(x,y,z)}n_T(E)\Delta E n_S^{\alpha}(E)\Delta E[1+P_T(E)P_S^{\alpha}(E)cos\varphi_{\alpha}(E)].
\end{equation}
This means that by multiplying the $LDOS$ with $\Delta E$ results in a dimensionless quantity, which is multiplied by the
conductance quantum $e^2/h$ in order to arrive at our $dI/dU$ expression.
Note that $n_T(E)\Delta E$ electron states from tip and $n_S^{\alpha}(E)\Delta E$ states from each surface atom
contribute to the differential current at energy $E$, and in our model, $dI/dU$ is proportional to the $LDOS$, which contains
both surface and tip electronic information.
If the two subsystems are calculated separately, it is possible to combine different levels of electronic structure
for tip and surface, see also Table \ref{Table_axis}, or include simplified model tip electronic structures into our approach.
For example, assuming an electronically flat maximally spin-polarized ($P_T(E)=1$) ideal magnetic tip with
e.g.\ $n_T(E)\Delta E=1$, the differential conductance reads
\begin{equation}
\label{Eq_dIdV-idealtip}
\frac{dI}{dU}(x,y,z,E)=\frac{e^2}{h}\sum_{\alpha}e^{-2\kappa(E)d_{\alpha}(x,y,z)}n_S^{\alpha}(E)\Delta E[1+P_S^{\alpha}(E)cos\varphi_{\alpha}(E)].
\end{equation}
Here, Eq.(\ref{Eq_kappa_TH}) has been assumed for the vacuum decay, and there is no $V$-dependence.

By measuring the energy with respect to the sample Fermi level as $E=E_F^S+eU$,
the energy dependence can be transformed to bias dependence $U$, as
\begin{eqnarray}
\label{Eq_dIdV-idealtip-V}
&&\frac{dI}{dU}(x,y,z,U)=\frac{e^2}{h}\sum_{\alpha}e^{-2\kappa(E_F^S+eU)d_{\alpha}(x,y,z)}\\
&\times&n_S^{\alpha}(E_F^S+eU)\Delta E[1+P_S^{\alpha}(E_F^S+eU)cos\varphi_{\alpha}(E_F^S+eU)].\nonumber
\end{eqnarray}
Similarly, the more general differential conductance, Eq.(\ref{Eq_dIdV}), can be recast as
\begin{eqnarray}
\label{Eq_dIdV-V}
&&\frac{dI}{dU}(x,y,z,U,V)=\frac{e^2}{h}(\Delta E)^2\sum_{\alpha}e^{-2\kappa(E_F^S+eU,V)d_{\alpha}(x,y,z)}\\
&\times&n_T(E_F^T+eU-eV)n_S^{\alpha}(E_F^S+eU)[1+P_T(E_F^T+eU-eV)P_S^{\alpha}(E_F^S+eU)cos\varphi_{\alpha}(E_F^S+eU)],\nonumber
\end{eqnarray}
where we used the fact that the tip Fermi level is shifted by $eV$ with respect to the sample Fermi level,
i.e.\ $E_F^T=E_F^S+eV$, and therefore $E=E_F^T+eU-eV$.

Virtual differential conductances have to be determined at $E_i$ points in a fine energy grid with $\Delta E$ resolution
within an energy window $[E_1(V,T),E_2(V,T)]$ corresponding to the applied bias voltage ($V$) and temperature ($T$).
A value of $10^{-3}$ eV has been used for $\Delta E$ in our calculations, while we tested a finer grid
($\Delta E=10^{-4}$ eV) as well, with no improvement of our results.
Finally, the tunneling current can be determined by the following energy integral,
\begin{eqnarray}
\label{Eq_Curr}
&&I(x,y,z,V,T)=\int\limits_{E_1(V,T)}^{E_2(V,T)}\frac{dE}{e}\frac{dI}{dU}(x,y,z,E,V)=\frac{\Delta E}{e}\sum_{E_1<E_i<E_2}\frac{dI}{dU}(x,y,z,E_i,V)=\\
&&\frac{e}{h}(\Delta E)^3\sum_{E_1<E_i<E_2}\sum_{\alpha}e^{-2\kappa(E_i,V)d_{\alpha}(x,y,z)}n_T(E_i)n_S^{\alpha}(E_i)[1+P_T(E_i)P_S^{\alpha}(E_i)cos\varphi_{\alpha}(E_i)],\nonumber
\end{eqnarray}
where the energy window is defined as
\begin{eqnarray}
\label{Eq_E1}
E_1(V,T)&=&\min\left(E_F^S,E_F^S+eV\right)-ln\left(3+\sqrt{8}\right)k_B T,\\
E_2(V,T)&=&\max\left(E_F^S,E_F^S+eV\right)+ln\left(3+\sqrt{8}\right)k_B T.
\label{Eq_E2}
\end{eqnarray}
Here, $E_F^S$ is the Fermi energy of the sample surface, and $V$ is the applied bias voltage. Broadening of electron states at
finite temperatures is considered according to Eqs.(\ref{Eq_PDOS}), (\ref{Eq_CDOS_noncoll}), (\ref{Eq_MDOS_noncoll}), and the
temperature dependent terms in the integral limits are the full width at half maximum of the energy-derivative of the
Fermi distribution function divided by 2, and $k_B$ is the Boltzmann constant.
Another, more precise way to include thermal effects in calculating the tunneling current is given in the Appendix of
Ref.\ \cite{passoni07} based on the Sommerfeld expansion, which can also be incorporated into our approach.
Lattice vibrations at nonzero temperatures are not taken into account.

From the obtained three-dimensional tunneling current maps, data can be extracted which are directly comparable to
experiments. In particular, current values can be shown in arbitrary $z=Z_C=const.$ planes or constant-value surface contours
can be defined. The first option corresponds to the constant height mode, $I(x,y,Z_C=const.,V,T)$, while the
second to the constant current mode of SP-STM, $I(x,y,z,V,T)=I_C=const.$ From the latter, a two-dimensional tip position map,
called the height profile, $z(x,y,V,T,I_C)$ can be extracted using logarithmic interpolation between grid points
$z_1<z_2$, if $I(x,y,z_1,V,T)>I_C>I(x,y,z_2,V,T)$, in the following way,
\begin{equation}
\label{Eq_Corrugation1}
z(x,y,V,T,I_C)=z_1+\Delta z\frac{ln(I_C)-ln(I(x,y,z_1,V,T))}{ln(I(x,y,z_2,V,T))-ln(I(x,y,z_1,V,T))},
\end{equation}
where $\Delta z=z_2-z_1=z_{i+1}-z_i=0.0529177\AA$ (0.1 a.u.) has been used in all calculations.
Alternatively, if $I_C$ has such a value which is not contained in the considered finite box above the surface,
i.e.\ if $I_C<I(z_{max})$, then
\begin{equation}
\label{Eq_Corrugation2}
z(x,y,V,T,I_C)=z_{max}+\Delta z\frac{ln(I_C)-ln(I(x,y,z_{max},V,T))}{ln(I(x,y,z_{max},V,T))-ln(I(x,y,z_{max-1},V,T))}.
\end{equation}
Surface corrugation can be determined from this $z(x,y)$ map. Note that the total current contains both topographic and magnetic
contributions, therefore $z(x,y)$ will be the simulated SP-STM image \cite{heinze06},
in our model we consider its bias and temperature dependence as well.
In periodic magnetic systems the magnetic unit cell can be identified in the simulated image \cite{wortmann01}.
Moreover, two different types of magnetic contrast can be defined. The first one is the apparent height difference of a
particular atom at $(x_i,y_i)$ lateral position imaged with a magnetic tip with parallel (P) and antiparallel (AP)
relative magnetic orientation, on the same constant current contour,
\begin{equation}
\label{Eq_Contrast1}
\Delta z_{i}(V,T,I_C)=z(x_i,y_i,V,T,I^{P}=I_C)-z(x_i,y_i,V,T,I^{AP}=I_C).
\end{equation}
The other magnetic contrast is the apparent height difference of two magnetic atoms at different lateral positions
$(x_i,y_i)$ and $(x_j,y_j)$ imaged with a fixed tip magnetization direction,
\begin{equation}
\label{Eq_Contrast2}
\Delta z_{ij}(V,T,I_C)=z(x_j,y_j,V,T,I=I_C)-z(x_i,y_i,V,T,I=I_C).
\end{equation}
This means apparent height differences of different surface atoms on the same SP-STM image, similarly as considered
e.g.\ for oppositely magnetized islands in Ref.\ \cite{hofer08tipH}.
Note that both magnetic contrasts depend on the bias voltage, temperature and the constant current value. This latter means,
in effect, dependence on the tip-sample distance.

The obtained tunneling current can also be decomposed into a non-spin-polarized (TOPO) and a spin-polarized (MAGN) part,
\begin{equation}
I_{TOTAL}(x,y,z,V,T)=I_{TOPO}(x,y,z,V,T)+I_{MAGN}(x,y,z,V,T),
\label{Eq_Curr_decomp}
\end{equation}
with
\begin{eqnarray}
\label{Eq_Curr_TOPO}
I_{TOPO}(x,y,z,V,T)&=&\frac{e}{h}(\Delta E)^3\sum_{E_1<E_i<E_2}\sum_{\alpha}e^{-2\kappa(E_i,V)d_{\alpha}(x,y,z)}n_T(E_i)n_S^{\alpha}(E_i),\\
I_{MAGN}(x,y,z,V,T)&=&\frac{e}{h}(\Delta E)^3\sum_{E_1<E_i<E_2}\sum_{\alpha}e^{-2\kappa(E_i,V)d_{\alpha}(x,y,z)}m_T(E_i)m_S^{\alpha}(E_i)cos\varphi_{\alpha}(E_i).
\label{Eq_Curr_MAGN}
\end{eqnarray}
$I_{TOPO}$ and $I_{MAGN}$ can be analyzed separately using the same way as described for the total current and they can be
related to SP-STM experiments using the differential magnetic mode \cite{wulfhekel10review}.
From the non-magnetic height profile, $z(x,y,V,T,I_{TOPO}=const.)$, the surface topography can be calculated, and
in periodic systems the chemical unit cell is revealed in the simulated image.

It has to be noted that the presented method can also be applied to study nonmagnetic systems, where all magnetic
contributions are equal to zero and the corresponding topographic STM images can be simulated.

Finally, it is important to note that following Ref.\ \cite{passoni07}, the real physical differential conductance measured in
experiments can be obtained as the derivative of the tunneling current, Eq.(\ref{Eq_Curr}), with respect to the bias voltage.
It can be related to our virtual differential conductance $dI/dU$ defined in Eq.(\ref{Eq_dIdV-V}) in the following way,
\begin{equation}
\label{Eq_dIdV-proper}
\frac{dI}{dV}(x,y,z,V',T=0)=\frac{dI}{dU}(x,y,z,V',V')+\int_{0}^{V'}dU\left.\frac{\partial}{\partial V}\frac{dI}{dU}(x,y,z,U,V)\right|_{V=V'}.
\end{equation}
Here, $T=0$ K temperature is considered for the reason of simplicity in the integral limits.
In the case of assuming an ideal magnetic tip, i.e.\ if $n_T(E)=const.$, $\overline{m}_T(E)=const.$, and $\kappa(E)$ has no
$V$-dependence as defined in Eq.(\ref{Eq_kappa_TH}), then $dI/dU$ has no $V$-dependence as in
Eq.(\ref{Eq_dIdV-idealtip-V}), and consequently, the integral term is zero. In that highly idealized setup,
$dI/dV(V')=dI/dU(V')$, such that Eq.(\ref{Eq_dIdV-idealtip-V}) is the real differential conductance with $U=V=V'$.
Moreover, $dI/dU$ can also be written as a sum of TOPO and MAGN parts,
\begin{equation}
\label{Eq_dIdV_decomp}
\frac{dI_{TOTAL}}{dU}(x,y,z,U)=\frac{dI_{TOPO}}{dU}(x,y,z,U)+\frac{dI_{MAGN}}{dU}(x,y,z,U),
\end{equation}
with
\begin{eqnarray}
\label{Eq_dIdV_TOPO}
\frac{dI_{TOPO}}{dU}(x,y,z,U)&=&\frac{e^2}{h}\Delta E\sum_{\alpha}e^{-2\kappa(E_F^S+eU)d_{\alpha}(x,y,z)}n_S^{\alpha}(E_F^S+eU),\\
\frac{dI_{MAGN}}{dU}(x,y,z,U)&=&\frac{e^2}{h}\Delta E\sum_{\alpha}e^{-2\kappa(E_F^S+eU)d_{\alpha}(x,y,z)}m_S^{\alpha}(E_F^S+eU)cos\varphi_{\alpha}(E_F^S+eU)],
\label{Eq_dIdV_MAGN}
\end{eqnarray}
and they can be analyzed separately. We return to the simulation of spin-polarized scanning tunneling spectroscopy (SP-STS)
based on the atom superposition method in the future.

\section{Results and Discussion}
\label{sec_res}

In order to demonstrate the capabilities of our model for simulating SP-STM on complex magnetic surfaces,
we consider a sample surface with noncollinear magnetic order. One ML Cr on Ag(111) is a prototype of
frustrated hexagonal antiferromagnets \cite{heinze06}. Due to the geometrical frustration of the
antiferromagnetic exchange interactions between Cr spin moments, its magnetic ground state has been
determined to be a noncollinear $120^{\circ}$ N\'eel state \cite{wortmann01}. We consider two possible N\'eel states with
opposite chiralities, which are energetically equivalent only in the absence of spin-orbit coupling.

We performed geometry relaxation and electronic structure calculations based on Density Functional Theory (DFT)
within the Generalized Gradient Approximation (GGA) implemented in the Vienna Ab-initio Simulation Package (VASP)
\cite{VASP2,VASP3,hafner08}. A plane wave basis set for electronic wavefunction expansion together with the
projector augmented wave (PAW) method \cite{kresse99} has been applied, while
the exchange-correlation functional is parametrized according to Perdew and Wang (PW91) \cite{pw91}.
For calculating the fully noncollinear electronic structure we used the VASP code as well \cite{hobbs00prb,hobbs00jpcm},
with spin-orbit coupling considered. This allows us to determine the N\'eel state with the energetically favored chirality.

We model the Cr/Ag(111) system by a slab of a five-layer Ag substrate and one-one monolayer Cr films on each side,
where the surface Cr layers
and the first Ag layers underneath have been fully relaxed. After relaxation the Cr-Ag interlayer distance is reduced by 9.5\%,
while the underneath Ag-Ag increased by 0.5\% compared to bulk Ag. A separating vacuum region of 14.6 $\AA$ width in the
surface normal ($z$) direction has been set up between neighboring supercell slabs.
The average electron workfunction above the Cr is calculated to be $\phi_S=4.47$ eV using Eq.(\ref{Eq_WF_average}).
We used an $11\times 11\times 1$ Monkhorst-Pack (MP) \cite{monkhorst} k-point grid for calculating the projected electron DOS
onto the surface Cr atoms in our ($\sqrt{3}\times\sqrt{3}$) magnetic surface unit cell.

\begin{figure*}
\includegraphics[width=1.0\textwidth,angle=0]{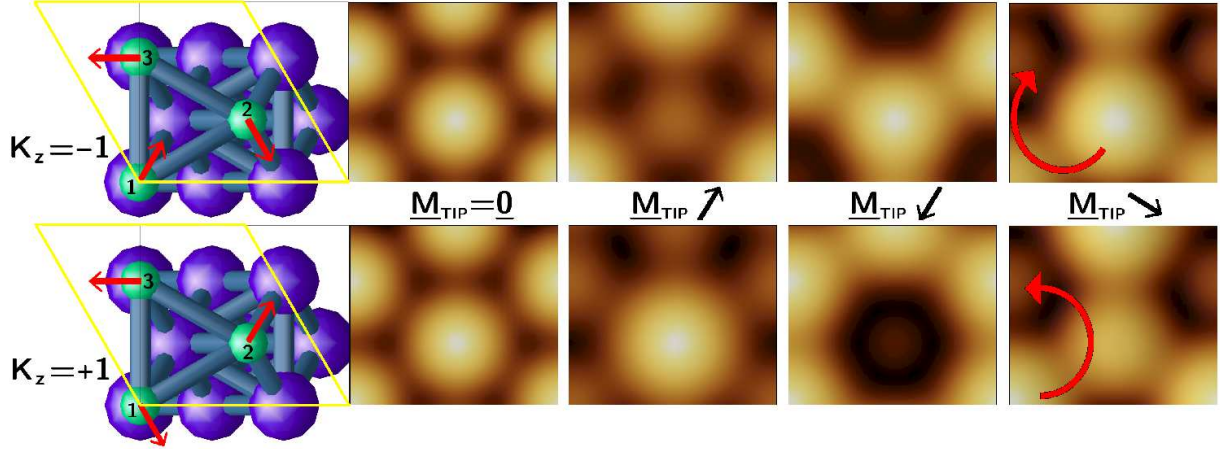}
\caption{\label{Fig1} (Color online) Surface geometry of 1 ML Cr on Ag(111) and simulated SP-STM images at 0 V bias voltage
depending on the tip magnetization direction ($\underline{M}_{TIP}$) assuming an ideal electronically flat maximally spin-polarized
tip. The Cr and Ag atoms are denoted by spheres colored by green (medium gray) and purple (dark gray), respectively,
while the magnetic moments of individual Cr atoms are indicated by (red) arrows in the left part of the figure.
The Cr atoms are explicitly labeled corresponding to the calculated chirality vector in Eq.(\ref{Eq_chirality}).
In addition, the ($\sqrt{3}\times\sqrt{3}$) magnetic unit cell is drawn by yellow (light gray) color.
In the two rows noncollinear N\'eel states with opposite chiralities and corresponding SP-STM images are shown.
In the last column, the decreasing levels of Cr apparent heights are indicated by circular arrows.
}
\end{figure*}

Performing fully noncollinear electronic structure calculations we obtained convergence to two different magnetic N\'eel states.
The magnetic surface unit cell with the converged magnetic moment directions are shown in the left part of Figure \ref{Fig1}. 
Each of the two N\'eel states can be characterized by a chirality vector, defined as \cite{antal08}
\begin{equation}
\label{Eq_chirality}
\overline{K}=\frac{2}{3\sqrt{3}}\left(\overline{e}_S^1\times\overline{e}_S^2+\overline{e}_S^2\times\overline{e}_S^3+\overline{e}_S^3\times\overline{e}_S^1\right).
\end{equation}
Here $\overline{e}_S^{\alpha}$ denotes the local spin quantization unit vector of the $\alpha$th Cr atom.
It is defined from the local magnetic moment, $\overline{M}_S^{\alpha}=\int_{-\infty}^{E_F^S}dE\overline{m}_S^{\alpha}(E)$,
similarly as in Eq.(\ref{Eq_axis}), i.e.\ $\overline{e}_S^{\alpha}=\overline{M}_S^{\alpha}/|\overline{M}_S^{\alpha}|$.
The magnitude of the magnetic moments of the Cr surface atoms are 3.73 $\mu_B$, with a
very small out-of-plane component, which is neglected when defining the chirality vectors.
Thus, in the first row of Figure \ref{Fig1},
$\overline{e}_S^1=(1/2,\sqrt{3}/2,0)$, $\overline{e}_S^2=(1/2,-\sqrt{3}/2,0)$, and $\overline{e}_S^3=(-1,0,0)$.
This corresponds to the chirality vector $\overline{K}=(0,0,-1)$ or simply $K_z=-1$.
Similarly, in the second row of Figure \ref{Fig1},
$\overline{e}_S^1=(1/2,-\sqrt{3}/2,0)$, $\overline{e}_S^2=(1/2,\sqrt{3}/2,0)$, and $\overline{e}_S^3=(-1,0,0)$
correspond to $K_z=+1$.
Comparing total energies of the two states we find that $K_z=-1$ is energetically favored by 1.1 meV compared to $K_z=+1$.
This finding is consistent with the magnetic ground state found for a Cr trimer island on the Au(111) substrate in
Ref.\ \cite{antal08}, where it was also shown that the Dzyaloshinskii-Moriya interaction is responsible for determining the
ground state magnetic chirality.
Performing a collinear calculation with spin-orbit coupling considered, we obtain a ferromagnetic (FM) state with in-plane
Cr atomic magnetic moments of 3.76 $\mu_B$.
It turns out that this FM state is 1.04 eV higher in energy than the $K_z=-1$ N\'eel state. The energy difference
of 346 meV/(magnetic atom) in favor of the N\'eel state is in good agreement with results of Ref.\ \cite{wortmann01}.
The out-of-plane FM state is 1 meV higher in energy than the in-plane FM state with the same magnitude of magnetic moments.

Simulation of SP-STM images can be performed using Eqs.(\ref{Eq_dIdV}) and (\ref{Eq_Curr}) in two ways:\\
(1) According to Heinze \cite{heinze06}, having chemically equivalent surface atoms the spin structure plays a more dominant
role compared to the detailed electronic structure in determining the main features of an SP-STM image. Following this, we can
take the collinear electronic structure (COLL) obtained from the in-plane ferromagnetic calculation,
and set the spin structure to the corresponding N\'eel state.\\
(2) As a more precise way, we can take the noncollinear electronic structure (NONCOLL), and there is no need to
prescribe the spin structure as it is naturally included in the electronic structure data.\\
The first approach is computationally cheaper, and can be applied to simulate larger scale images \cite{heinze06}.
On the other hand, calculation of the noncollinear electronic structure is computationally more demanding but more realistic.

\begin{figure*}
\includegraphics[width=1.0\textwidth,angle=0]{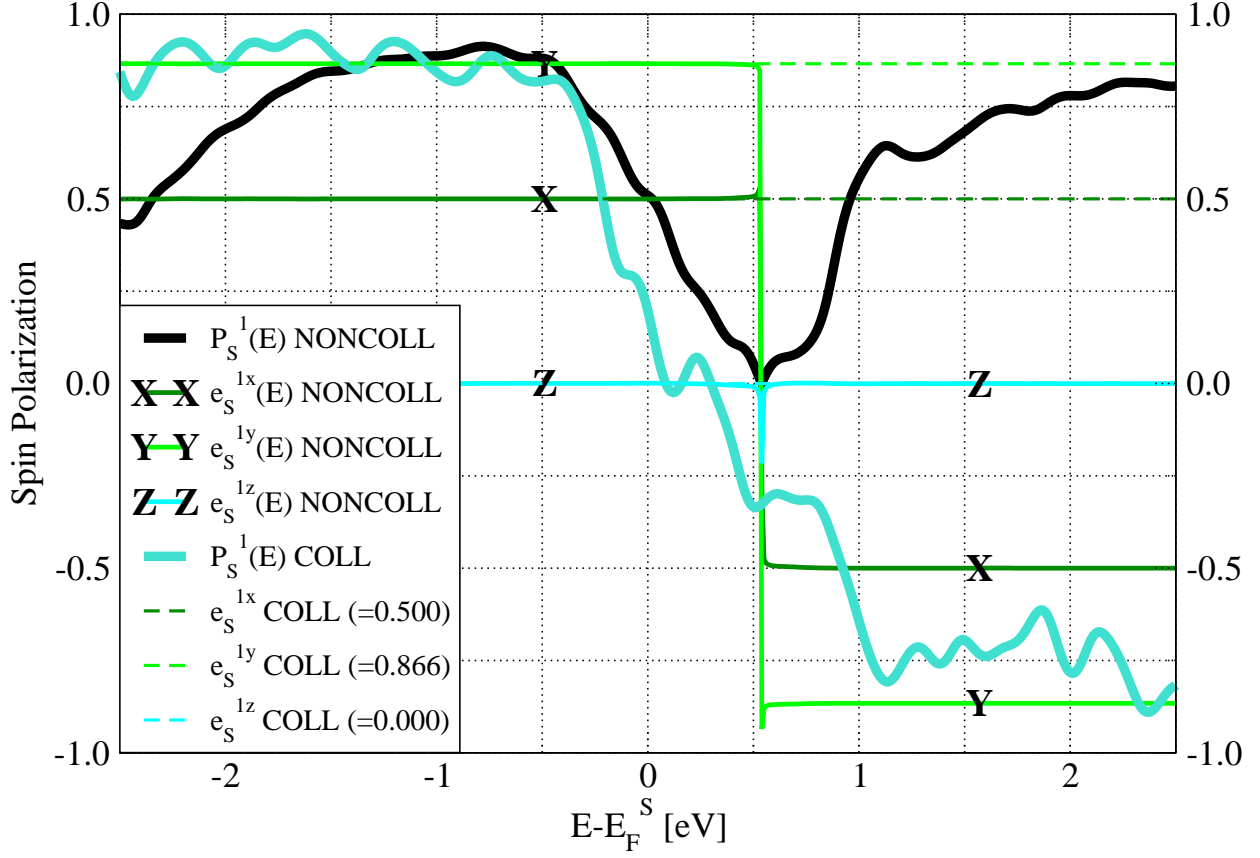}
\caption{\label{Fig2} (Color online) Comparison of energy dependent spin polarization vectors,
$\overline{P}_S^1(E)=P_S^1(E)(e_S^{1x}(E),e_S^{1y}(E),e_S^{1z}(E))$, of the surface Cr atom labeled by "1" in Figure \ref{Fig1}
in the N\'eel state with $K_z=-1$ chirality, calculated from noncollinear (NONCOLL) and collinear (COLL) electronic structure.
In the collinear case the local spin quantization axis is $\overline{e}_S^1=(1/2,\sqrt{3}/2,0)$ in the basis of ($e^x,e^y,e^z$),
and is independent of energy. Reversal of the spin polarization vector occurs at $P_S^1(E)$ values of opposite sign.
In the noncollinear case $P_S^1(E)$ is always positive due to Eq.(\ref{Eq_Spinpol_noncoll}) and the spin polarization vector
reversal is observed as the sign change of $e_S^{1x}(E)$ and $e_S^{1y}(E)$ at 0.54 V. Here, the
rotation direction of the spin polarization vector is through negative $e_S^{1z}(E)$ components.
}
\end{figure*}

In our Cr/Ag(111) system we calculated the tunneling current in a box above the magnetic unit cell
containing 153000 (34x30x150) grid points with a $0.15\AA$ lateral and $0.0529177\AA$ horizontal resolution.
Figure \ref{Fig1} shows simulated constant current SP-STM images for the two N\'eel states at zero bias voltage, assuming an
ideal electronically flat maximally spin-polarized tip based on Eq.(\ref{Eq_dIdV-idealtip}) with various magnetization directions
following the first method.
These are in qualitatively good agreement with previous simulations \cite{wortmann01,heinze06}. Using a nonmagnetic tip,
all surface Cr atoms appear to be of equal height (one height level), i.e.\ the surface topography is seen.
As the spin polarization of the Cr atoms at the Fermi energy is positive (see Figure \ref{Fig2}), and the
tip spin polarization is set to +1, the Cr atom with parallel/antiparallel magnetization direction relative to the tip appears
to be higher/lower than the other two Cr atoms, which have the same apparent height due to symmetry (two height levels).
Comparing the images, it is clear that a contrast reversal occurs when turning the tip magnetization to opposite direction.
This magnetic contrast can be quantified according to Eq.(\ref{Eq_Contrast1}).
By setting the tip magnetization direction perpendicular to a Cr magnetic moment, a structure with three height levels occurs.
This means that all Cr atoms in the magnetic unit cell have different apparent heights. This is due to the variation of
the angles between the local Cr magnetic moments and the tip magnetization, e.g.\ for $K_z=-1$,
$\varphi_1=90^{\circ},\varphi_2=30^{\circ},$ and $\varphi_3=150^{\circ}$.
Determining the chirality of the magnetic structure from experimental SP-STM images is only possible in such a scenario if
the tip magnetization direction is not parallel with the magnetic moment of any of the surface atoms.
In our example of the Cr/Ag(111) system the three apparent height levels follow a different order in the magnetic unit cell
corresponding to the different chiralities.
The decreasing levels of Cr apparent heights are indicated by circular arrows in the last column of Figure \ref{Fig1}.
Apparent height differences of individual atoms on the same image define another kind of magnetic contrast, see
Eq.(\ref{Eq_Contrast2}).
Generally, the determining factor for the apparent height of magnetic atoms in zero bias ($V$=0 V) measurements is
the effective spin polarization (ESP) at the common Fermi level, $P_T(E_F^S)P_S^{\alpha}(E_F^S)cos\varphi_{\alpha}(E_F^S)$,
similarly as it was identified as the governing factor for the height of differential tunneling spectra at particular energies
\cite{palotas11sts}.
A positive ESP results in higher tunneling current at a fixed distance above a magnetic surface atom,
while the opposite holds for negative ESP. Considering a constant current contour, thus,
results in a higher apparent height for the atom with positive, while a lower height with negative ESP,
compared to the topographic heights.

Let us analyze the consequences of the choice of the collinear (COLL) or noncollinear (NONCOLL) electronic structure
for the SP-STM images in more detail.
Taking the noncollinear electronic structure we obtained qualitatively similar images at zero bias as shown in Figure \ref{Fig1},
thus, Heinze is right \cite{heinze06} with the quality of the SP-STM images calculated at the sample Fermi energy using either
COLL or NONCOLL electronic structure.
The different spin polarization value of the Cr atoms at the Fermi level, however, results in different magnetic contrasts.
According to Figure \ref{Fig2}, the spin polarization of the Cr atoms is 0.20 and 0.51 considering the COLL and NONCOLL
electronic structure, respectively. Consequently, we expect that the magnetic contrast is higher in the NONCOLL SP-STM image.
Indeed, e.g.\ taking a constant current contour of $10^{-4}$ nA at parallel tip magnetization direction to the surface Cr atom
labeled by "1" (Cr1), we find $\Delta z_{12}^{COLL}(V=0$V$,T=4.2$K$,I_C=10^{-4}$nA$)=0.07$ $\AA$ and
$\Delta z_{12}^{NONCOLL}(V=0$V$,T=4.2$K$,I_C=10^{-4}$nA$)=0.21$ $\AA$ magnetic contrasts for COLL and NONCOLL images, respectively.
Moreover, the $10^{-4}$ nA contour is closer to the sample surface in the NONCOLL case.
It is worth to compare magnetic contrasts of COLL and NONCOLL images on constant current contours having the same apparent
height for Cr1. For example, the Cr1 apparent height of 3.35 $\AA$ is obtained at $10^{-4}$ nA in the COLL and
$5\times 10^{-5}$ nA in the NONCOLL image. The magnetic contrast in the new contour of the NONCOLL image is
$\Delta z_{12}^{NONCOLL}(V=0$V$,T=4.2$K$,I_C=5\times 10^{-5}$nA$)=0.18$ $\AA$.
Thus, we find that the magnetic contrast ratio of NONCOLL and COLL images at the same Cr1 apparent height of 3.35 $\AA$,
$0.18\AA /0.07\AA$ equals to the spin polarization ratio of Cr1 NONCOLL and COLL electronic structures at the Fermi level,
i.e.\ $0.51/0.20$.

In the following we consider the magnetic N\'eel state with $K_z=-1$ chirality since it has been identified as the ground state.
Figure \ref{Fig2} compares the energy dependent spin polarization vectors of Cr1 in Figure \ref{Fig1},
calculated from NONCOLL and COLL electronic structures.
The spin polarization vector is defined as $\overline{P}_S^1(E)=P_S^1(E)\overline{e}_S^1(E)$, where $P_S^1(E)$
is calculated using Eq.(\ref{Eq_Spinpol_coll}) and Eq.(\ref{Eq_Spinpol_noncoll}) in the COLL and NONCOLL case, respectively.
Taking the collinear electronic structure, the local spin quantization axis of Cr1 is set to the local magnetic moment direction
with neglecting the small out-of-plane component, $\overline{e}_S^1=(1/2,\sqrt{3}/2,0)$, and it is independent of energy.
Reversal of the spin polarization vector occurs at $P_S^1(E)$ values of opposite sign. Note that three sign changes occur in the
[0.0 eV,0.3 eV] energy interval with respect to the Fermi level, using an $11\times 11\times 1$ MP k-point grid. We tested a
denser $15\times 15\times 3$ MP k-point grid as well, resulting in a qualitatively similar spin polarization. For computational
and comparison reasons, we chose the $11\times 11\times 1$ MP k-point set for calculating the NONCOLL electronic structure.
While, in the noncollinear case $P_S^1(E)$ is always positive due to Eq.(\ref{Eq_Spinpol_noncoll}), one spin polarization vector
reversal is observed at 0.54 eV above the Fermi level. The indication for this reversal is the sign change of
$e_S^{1x}(E)$ and $e_S^{1y}(E)$, i.e.\ going away from the Fermi energy, the local spin quantization axis changes from
$\overline{e}_S^1\approx (1/2,\sqrt{3}/2,0)$ to $\overline{e}_S^1\approx (-1/2,-\sqrt{3}/2,0)$ at 0.54 eV.
Here, however, the $e_S^{1z}(E)$ components are not exactly zero, but they are in the order of $10^{-6}$ to $10^{-2}$ in
the whole energy range with the exception of $e_S^{1z}(0.54eV)=-0.21$. This latter value indicates that the rotation direction
of the spin polarization vector at 0.54 eV is through negative $e_S^{1z}$ components.
Since $\overline{e}_S^1(E)$ is a unit vector at all energies, the presence of small $e_S^{1z}(E)$ components also means that
the other vector components are $|e_S^{1x}(E)|\approx 1/2$ and $|e_S^{1y}(E)|\approx \sqrt{3}/2$.
By comparing the $P_S^1(E)$ spin polarization function of COLL and NONCOLL electronic structures, we can state qualitative
agreement.

\begin{figure*}
\includegraphics[width=1.0\textwidth,angle=0]{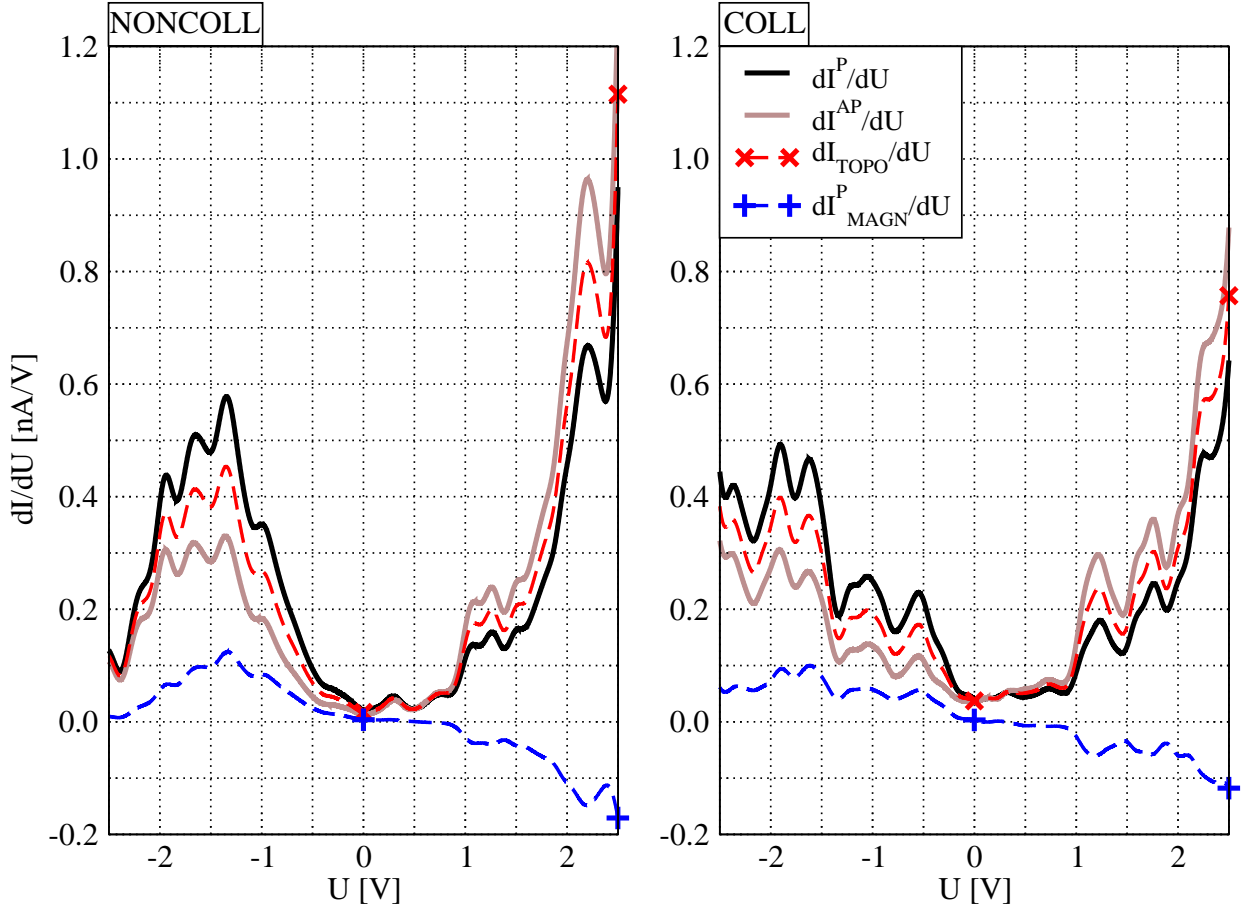}
\caption{\label{Fig3} (Color online) Simulated single point differential tunneling spectra $dI^P/dU$ and $dI^{AP}/dU$
3.5 $\AA$ above the surface Cr atom labeled by "1" in Figure \ref{Fig1} in the N\'eel state with $K_z=-1$ chirality,
assuming parallel (P) and antiparallel (AP) tip magnetization direction with respect to that of Cr1,
applying an ideal electronically flat maximally spin-polarized tip according to Eq.(\ref{Eq_dIdV-idealtip-V}).
Left and right parts correspond to spectra obtained from noncollinear (NONCOLL) and collinear (COLL) electronic structures
of the sample, respectively.
Topographic ($dI_{TOPO}/dU$) and magnetic ($dI^P_{MAGN}/dU=-dI^{AP}_{MAGN}/dU$) contributions are given according to
Eq.(\ref{Eq_dIdV_TOPO}) and Eq.(\ref{Eq_dIdV_MAGN}), respectively.
}
\end{figure*}

Let us compare simulated single point differential conductance spectra based on NONCOLL and COLL electronic structures.
Figure \ref{Fig3} shows such simulated spectra $z=3.5$ $\AA$ above the Cr1 atom in Figure \ref{Fig1} with assumed parallel (P) and
antiparallel (AP) tip magnetization direction using an ideal magnetic tip. We showed at the end of section \ref{sec_spstm} that
for the considered ideal magnetic tip $dI/dU$ (Eq.(\ref{Eq_dIdV-idealtip-V})) is the real differential conductance.
According to Eq.(\ref{Eq_dIdV_decomp}), the topographic and magnetic contributions can be calculated separately.
Determining $dI_{TOPO}/dU(z,U)$ (red dashed line with symbol "X") and $dI_{MAGN}^P/dU(z,U)$ (blue dashed line with symbol "+") is
sufficient to draw $dI^P/dU(z,U)$ (black solid line) and $dI^{AP}/dU(z,U)$ (brown (gray) solid line) since
\begin{eqnarray}
dI^P/dU(z,U)&=&dI_{TOPO}/dU(z,U)+dI_{MAGN}^P/dU(z,U)\nonumber\\
dI^{AP}/dU(z,U)&=&dI_{TOPO}/dU(z,U)-dI_{MAGN}^P/dU(z,U).
\end{eqnarray}
Here, we took into account that the magnetic contribution for the AP tip magnetization direction $dI_{MAGN}^{AP}/dU$ equals to
$-dI_{MAGN}^P/dU$, since $cos\varphi$ changes sign.
The COLL and NONCOLL spectra have slightly different peak positions due to the details of the electronic structure.
We find that $dI^P/dU>dI_{TOPO}/dU>dI^{AP}/dU$ below $U=0.54$ V, while $dI^P/dU<dI_{TOPO}/dU<dI^{AP}/dU$
above $U=0.54$ V, calculated by using NONCOLL electronic structure. The relation of these quantities is determined be the sign of
the magnetic contribution at the given bias, i.e.\ $dI_{MAGN}^{P}/dU(U<0.54$V$)>0$, and $dI_{MAGN}^{P}/dU(U>0.54$V$)<0$. 
On the other hand, for the COLL case, there are three sign changes of $dI_{MAGN}^{P}/dU$ at 85 mV, 160 mV, and 300 mV, resulting
in $dI^P/dU>dI_{TOPO}/dU>dI^{AP}/dU$ below $U=85$ mV and $dI^P/dU<dI_{TOPO}/dU<dI^{AP}/dU$ above $U=300$ mV. The magnetic
contribution is small between 85 mV and 300 mV and the difference between spectra is less than 0.02 nA/V in this bias range.

\begin{figure*}
\includegraphics[width=1.0\textwidth,angle=0]{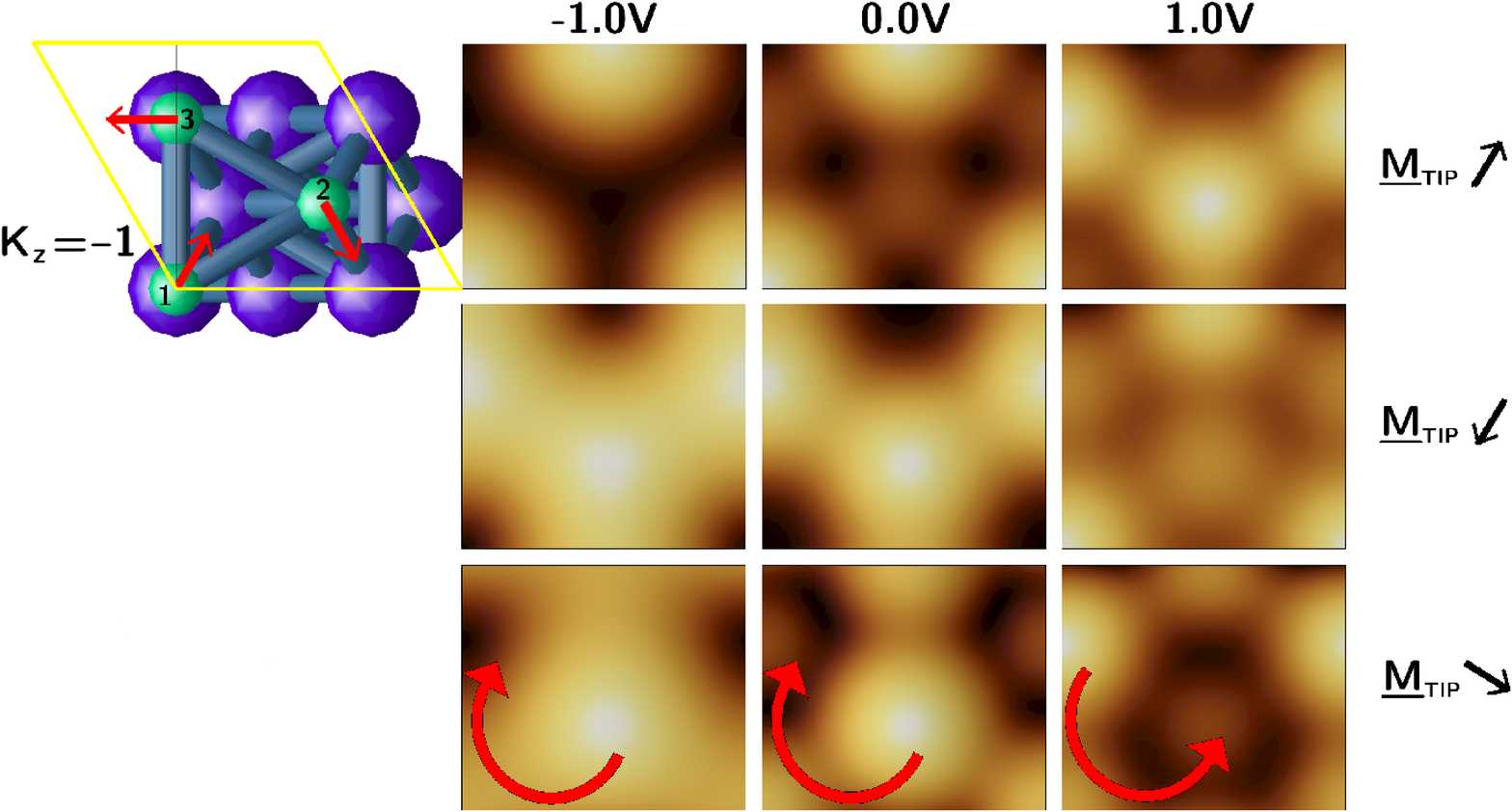}
\caption{\label{Fig4} (Color online) Simulated SP-STM images depending on the bias voltage and the tip magnetization direction
assuming an ideal electronically flat maximally spin-polarized tip. The magnetic contrast is reversed between 0.0 V and 1.0 V.
In the last row, the decreasing levels of Cr apparent heights are indicated by circular arrows.
The surface geometry of 1 ML Cr on Ag(111), its magnetic structure with $K_z=-1$ chirality, and the considered tip magnetization
directions are explicitly shown, similarly as in Figure \ref{Fig1}.
}
\end{figure*}

In the following we use the NONCOLL electronic structure for the Cr/Ag(111) sample surface.
By including energy dependent electronic structure of sample and tip into our model, we can study the bias dependent magnetic
contrast and its tip dependence as well. Figure \ref{Fig4} shows simulated SP-STM images for various tip magnetization directions
at -1 V, 0 V and +1 V bias voltages assuming an ideal magnetic tip. We find qualitatively similar images for -1 V and 0 V
for the corresponding tip magnetization direction. This means that the $\Delta z_{12}$ magnetic contrast between Cr1 and Cr2
has the same sign at -1 V and 0 V. However, $\Delta z_{12}$ at the same Cr1 apparent height increases at -1 V compared to 0 V
for all tip magnetization directions. This can be explained by the integrated $dI_{MAGN}^{P}$ contribution, which does not change
sign in this bias range, see blue dashed line with symbol "+" in the left part of Figure \ref{Fig3}.
On the other hand, the results show that the magnetic contrast is reversed at +1 V compared
to the other two studied bias voltages. This contrast reversal is observed for all tip magnetization directions.
It is interesting to find that on the image with three height levels the apparent heights change order in such a way that the
image at +1 V looks like that the N\'eel state would have an opposite chirality compared to 0 V or -1 V, see the indicated
circular arrows in the last row of Figure \ref{Fig4}. This finding highlights the importance of the applied bias voltage and
suggests that one has to be careful when interpreting the magnetic structure from experimentally observed SP-STM images.
Based on our theoretical study we can also conclude that the magnetic contrast reversal occurs between 0 V and +1 V bias voltages.
This contrast reversal is solely due to the sample electronic structure since the ideal magnetic tip is electronically featureless.

Dependence of the magnetic contrast on the tip electronic structure can be studied by considering different tip models.
As an example we chose a ferromagnetic Ni tip.
Such tips are routinely used in SP-STM and STS experiments \cite{heinrich09,heinrich10}.
The Ni tip has been modeled by a seven-layer Ni film slab with (110) orientation, having one-one Ni apex atoms on both surfaces,
i.e.\ with a double vacuum boundary. Here, the apex atom and the topmost surface layers have been relaxed on both sides.
The interaction between apex atoms in neighboring supercells is minimized by choosing a $3\times 3$ surface cell, and
a 15.4 $\AA$ wide separating vacuum region in $z$ direction.
Moreover, an $11\times 11\times 1$ MP k-point grid has been chosen for obtaining the projected DOS onto the apex atom.
The electronic structure of the apex is given in the top part of Figure 1 of Ref.\ \cite{palotas11sts}. We obtain a spin
polarization of $P_T=-0.91$ at the Fermi level, $E_F^T$, and $|P_T(E)|>0.8$ between $E_F^T-0.3$eV and $E_F^T+0.3$eV.
Employing Eq.(\ref{Eq_WF_local}), the local electron workfunction above the tip apex is $\phi_T=4.52$ eV,
and Eq.(\ref{Eq_kappa_WKB}) has been used to determine the vacuum decay.

\begin{figure*}
\includegraphics[width=1.0\textwidth,angle=0]{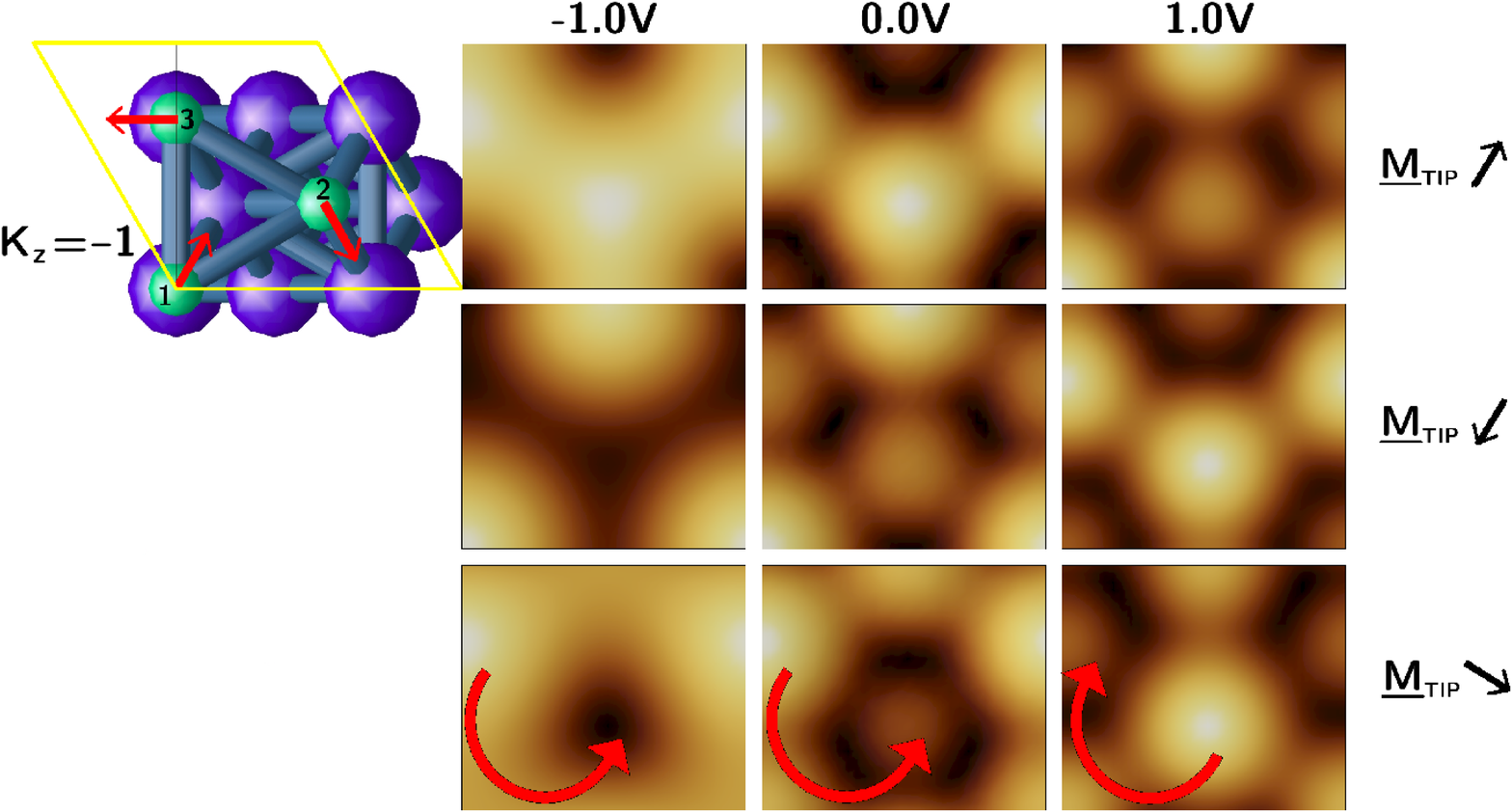}
\caption{\label{Fig5} (Color online) Simulated SP-STM images depending on the bias voltage and the tip magnetization direction
assuming a model Ni tip. The magnetic contrast is reversed compared to images obtained by using the ideal magnetic tip
(compare to Figure \ref{Fig4}), and there is a bias dependent contrast reversal between 0.0 V and 1.0 V.
In the last row, the decreasing levels of Cr apparent heights are indicated by circular arrows.
The surface geometry of 1 ML Cr on Ag(111), its magnetic structure with $K_z=-1$ chirality, and the considered tip magnetization
directions are explicitly shown, similarly as in Figure \ref{Fig1}.
}
\end{figure*}

Figure \ref{Fig5} shows simulated SP-STM images for various tip magnetization directions at -1 V, 0 V and +1 V bias voltages
including the electronic structure of the Ni tip into our model. By comparing images to those shown in Figure \ref{Fig4} obtained
by using an ideal magnetic tip, we find that the magnetic contrast is the opposite for each picture. This is due to the negative
spin polarization (-0.91) of the Ni tip apex at its Fermi level \cite{palotas11sts}. Note that the spin polarization of the ideal
magnetic tip was assumed to be +1 in the whole energy range. Similarly as in Figure \ref{Fig4}, we find qualitatively similar
images for -1 V and 0 V for the corresponding tip magnetization directions with higher magnetic contrast at -1 V compared to 0 V.
Again, the magnetic contrast is reversed at +1 V compared to the other two studied bias voltages.
This effect is highlighted in the last row of Figure \ref{Fig5} showing the decreasing levels of Cr apparent heights by circular
arrows, thus, indicating a bias dependent apparent magnetic chirality.
The results suggest that different tips can completely reverse the magnetic contrast.
This effect has to be taken into account when determining the magnetic structure from experimentally observed SP-STM images.

\section{Conclusions}
\label{sec_conc}

We extended the atom-superposition-based method of Heinze \cite{heinze06} for simulating spin-polarized scanning tunneling
microscopy by including the tip electronic structure, bias voltage, and the capability of incorporating the
fully noncollinear electronic structure. Taking the tip electronic structure into account, the effect of a richer variety
of electronic structure properties can be investigated on the tunneling transport within the indicated approximations
(atom superposition, spherical vacuum decay).
The method is computationally cheap and it can be applied based on results of any ab initio electronic structure code.
Taking the prototype frustrated hexagonal antiferromagnetic system, Cr monolayer on Ag(111) in a noncollinear magnetic
$120^{\circ}$ N\'eel state, we determined its ground state magnetic chirality and simulated SP-STM images at different bias
voltages to illustrate the applicability of our method. We related the magnetic contrast of the zero bias images
to the effective spin polarization at the sample Fermi level. Moreover, we illustrated the importance of the energy dependent
local spin quantization axes by comparing collinear and noncollinear electronic structure of a particular surface Cr atom and its
effect on single point tunneling spectra. Finally, we showed evidence that the magnetic contrast is sensitive to the
tip electronic structure, and this contrast can be reversed depending on the bias voltage.

\section{Acknowledgments}

Financial support of the Magyary Foundation, EEA and Norway Grants, the Hungarian Scientific Research Fund (OTKA PD83353, K77771),
the Bolyai Research Grant,
and the New Sz\'echenyi Plan of Hungary (Project ID: T\'AMOP-4.2.1/B-09/1/KMR-2010-0002) is gratefully acknowledged.

\begin{table}[b]
\caption{\label{Table_axis}
Combinations of taking into account energy dependence (Y) or independence (N) of the local spin quantization axes of surface atoms
and the tip apex, and consequence for the energy dependence of the angle $\varphi_{\alpha}$ between the surface local and the tip
quantization axes. The combinations framed by black solid lines are considered in the present work, while the one with the
gray dashed line corresponds to the studied system in Ref.\ \cite{palotas11sts}.}
\includegraphics[width=0.7\textwidth,angle=0]{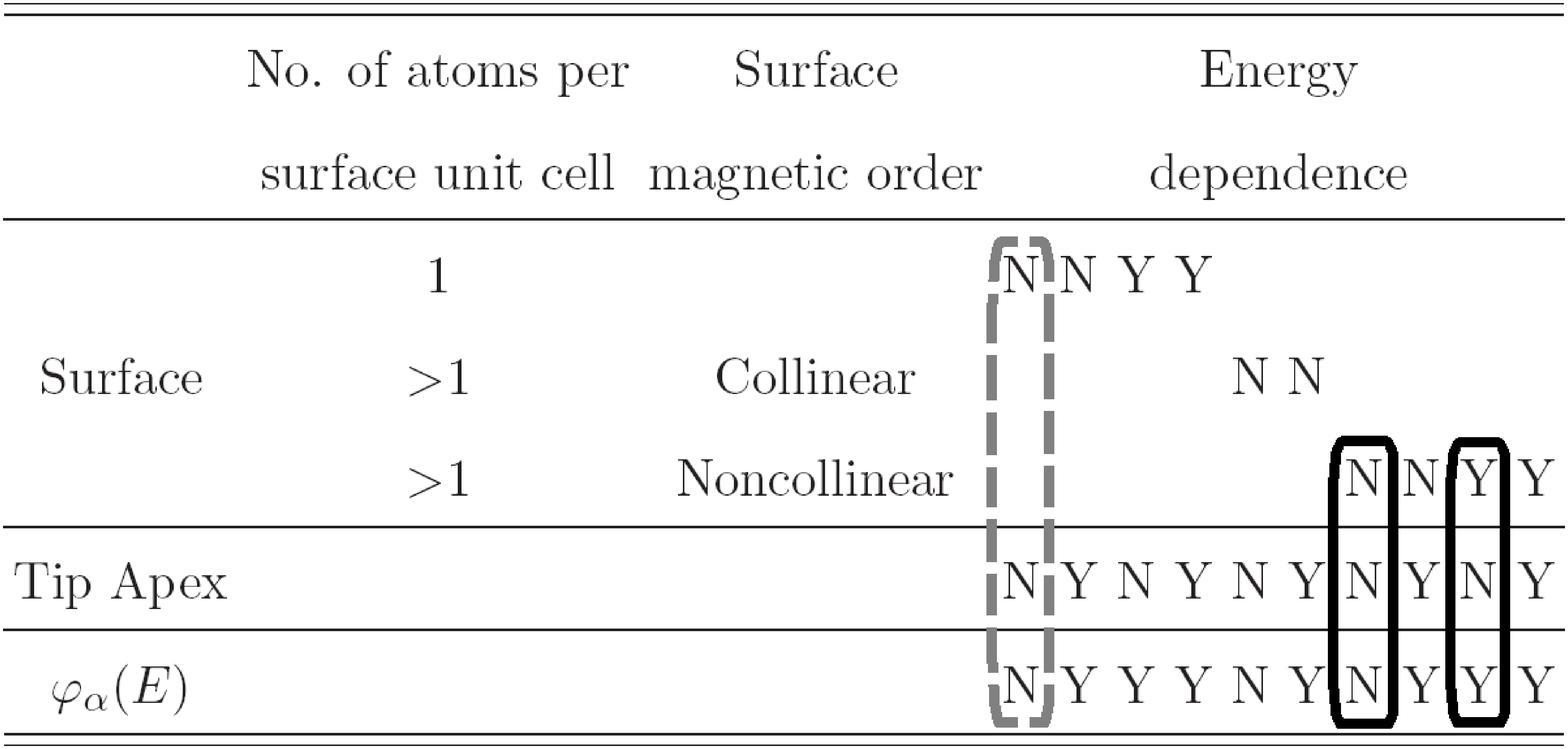}
\end{table}

\end{document}